\documentclass[superscriptaddress, nofootinbib, prd]{revtex4}[12pt]
\usepackage{graphicx}
\graphicspath{{figs/}}
\usepackage {amsmath, amssymb, rotating}
\usepackage{hyperref}
\usepackage[usenames,dvipsnames]{color}
\usepackage{mdwlist} 
\usepackage{slashed}
\usepackage{verbatim}
\usepackage{mathrsfs}

\newcommand{\vev}{\langle h \rangle }
\newcommand{\Mpl}{M_\text{pl}}
\newcommand{\Hi}{H_\text{i}}
\newcommand{\GeV}{{\rm\; GeV}}
\newcommand{\TeV}{{\rm\; TeV}}

\begin{document}

\title{Cosmological Relaxation of the Electroweak Scale}

\author{Peter W. Graham}
\affiliation{Stanford Institute for Theoretical Physics, Department of Physics, Stanford University, Stanford, CA 94305}

\author{David 
E.~Kaplan}
\affiliation{Stanford Institute for Theoretical Physics, Department of Physics, Stanford University, Stanford, CA 94305}
\affiliation{Department of Physics \& Astronomy, The Johns Hopkins University, Baltimore, MD  21218}
\affiliation{Berkeley Center for Theoretical Physics, Department of Physics, University of California, Berkeley, CA 94720}
\affiliation{Kavli Institute for the Physics and Mathematics of the Universe (WPI), Todai Institutes for Advanced Study, University of Tokyo, Kashiwa 277-8583, Japan}

\author{Surjeet Rajendran}
\affiliation{Berkeley Center for Theoretical Physics, Department of Physics, University of California, Berkeley, CA 94720}

\begin{abstract}
A new class of solutions to the electroweak hierarchy problem is presented that does not require either weak scale dynamics or anthropics.
Dynamical evolution during the early universe drives the Higgs mass to a value much smaller than the cutoff.
The simplest model has the particle content of the standard model plus a QCD axion and an inflation sector. 
The highest cutoff achieved in any technically natural model is $10^8$ GeV.
\end{abstract}

\date{\today}
\maketitle


\section{Introduction}


In the 1970's, Wilson \cite{Wilson} had discovered that a fine-tuning seemed to be required of any field theory which completed the standard model Higgs sector, unless its new dynamics appeared at the scale of the Higgs mass.  Since then, there have essentially been one and a half explanations proposed: dynamics and anthropics.

Dynamical solutions propose new physics at the electroweak scale which cuts off contributions to the  quadratic term in the Higgs potential.
Proposals include supersymmetry, compositeness for the Higgs (and its holographic dual), extra-dimensions, or even quantum gravity at the electroweak scale \cite{Dimopoulos:1981zb, Randall:1999ee, ArkaniHamed:1998rs, Susskind:1978ms, Weinberg:1979bn}.
While these scenarios all lead to a technically natural electroweak scale, collider and indirect constraints force these models into fine-tuned regions of their parameter spaces.  Anthropics, on the other hand, allows for the tuning, but assumes the existence of a multiverse.  Its difficulty is in the inherent ambiguity in defining both probability distributions and observers.

We propose a new class of solutions to the hierarchy problem.  The Lagrangian of these models are not tuned, and yet have no new physics at the weak scale cutting off loops.  In fact, the simplest model has no new physics at the weak scale at all.
It is instead dynamical evolution of the Higgs mass in the early universe that chooses an electroweak scale  parametrically smaller than the cutoff of the theory.  Our theories take advantage of the simple fact that the Higgs mass-squared equal to zero, while not a special point in terms of symmetries, {\it is} a special point in terms of dynamics, namely it is the point where the weak force spontaneously breaks and the theory enters a different phase\footnote{Of course QCD actually already breaks electroweak symmetry but at a much smaller scale.  In addition, there is no phase transition between broken and unbroken electroweak symmetry in the known standard model.  However, since both of these statements are approximately true, we will continue to use these terms in the text for brevity.}.  It is this which chooses the weak scale, allowing it to be very close to zero.  This mechanism takes some inspiration from  Abbott's attempt to solve the cosmological constant problem \cite{Abbott:1984qf}.

Our models only make the weak scale technically natural \cite{'tHooft:1980xb} and we have not yet attempted to UV complete them for a fully natural theory though there are promising directions \cite{Silverstein:2008sg, McAllister:2008hb, Kaloper:2008fb, Kaloper:2011jz, delaFuente:2014aca}.  Note, technical naturalness means a theory still contains small parameters, yet they are quantum mechanically stable and therefore one can imagine field-theoretic UV completions.
In addition, our models require large field excursions, far above the cutoff, and small couplings.
We judge the success of our models by how far they are able to naturally raise the cutoff of the Higgs.  
Our simplest model can raise the cutoff to $\sim 1000$ TeV, 
and we present a second model which can raise the cutoff up to $\sim 10^5$ TeV.


\section{Minimal Model}
\label{sec: QCD}


In our simplest model, the particle content below the cutoff is just the standard model plus the QCD axion \cite{Peccei:1977hh,Weinberg:1977ma,Wilczek:1977pj}, with an unspecified inflation sector.  Of course, by itself the QCD axion does not solve the hierarchy problem.  The only changes we need to make to the normal axion model are to give the axion a very large (non-compact) field range, and a soft symmetry-breaking coupling to the Higgs.

The axion will have its usual periodic potential, but now extending over many periods for a total field range that is parametrically larger than the cutoff (and may be larger than the Planck scale), similar to recent inflation models such as axion monodromy \cite{Silverstein:2008sg, McAllister:2008hb, Kaloper:2008fb, Kaloper:2011jz, delaFuente:2014aca}.  The exact (discrete) shift-symmetry of the axion potential is then softly broken by a small dimensionful coupling to the Higgs.  This small coupling will help set the weak scale, and will be technically natural, making the weak scale technically natural and solving the hierarchy problem.

We add to the standard model Lagrangian the following terms:
\begin{equation}
(-M^2 + g \phi) |h|^2 + V(g \phi ) + \frac{1}{32 \pi^2}\frac{\phi}{f}{\tilde G}^{\mu\nu}G_{\mu\nu} 
\end{equation}
where $M$ is the cutoff of the theory (where SM loops are cutoff), $h$ is the Higgs doublet, $G_{\mu\nu}$ is the QCD field strength (and ${\tilde G}^{\mu\nu}=\epsilon^{\mu\nu\alpha\beta}G_{\alpha\beta}$), $g$ is our dimensionful coupling, and we have neglected order one numbers.  We have set the mass of the Higgs to be at the cutoff $M$ so that it is natural.
The field $\phi$ is like the QCD axion, but can take on field values much larger than $f$.  However, despite its non-compact nature it has all the properties of the QCD axion with couplings set by $f$.
Setting $g\rightarrow 0$, the Lagrangian has a shift symmetry $\phi \rightarrow \phi + 2\pi f$ (broken from a continuous shift symmetry by non-perturbative QCD effects).  Thus, $g$ can be treated as a spurion that breaks this symmetry entirely.  This coupling can generate small potential terms for $\phi$, and we take the potential with technically natural values by expanding in powers of $g \phi$.  Non-perturbative effects of QCD produce an additional potential for $\phi$, satisfying the discrete shift symmetry.  Below the QCD scale, our potential becomes
\begin{equation}
(-M^2 + g \phi) |h|^2 + \left( g M^2 \phi + g^2 \phi^2 + \cdots \right) + \Lambda^4 \cos(\phi/f) 
\end{equation}
where the ellipsis represents terms higher order in $g \phi / M^2$, and thus we take the range of validity for $\phi$ in this effective field theory to be $\phi \lesssim M^2/g$.   We have approximated the periodic potential generated by QCD as a cosine, but in fact the precise form will not affect our results.  Of course $\Lambda$ is very roughly set by QCD, but with important corrections that we discuss below.  Both $g$ and $\Lambda$ break symmetries and it is technically natural for them to be much smaller than the cutoff.  The parameters $g$ and $\Lambda$ are responsible for the smallness of the weak scale.  This model plus inflation solves the hierarchy problem.

\begin{figure}[htbp]
\begin{center}
\includegraphics[width=0.75 \textwidth]{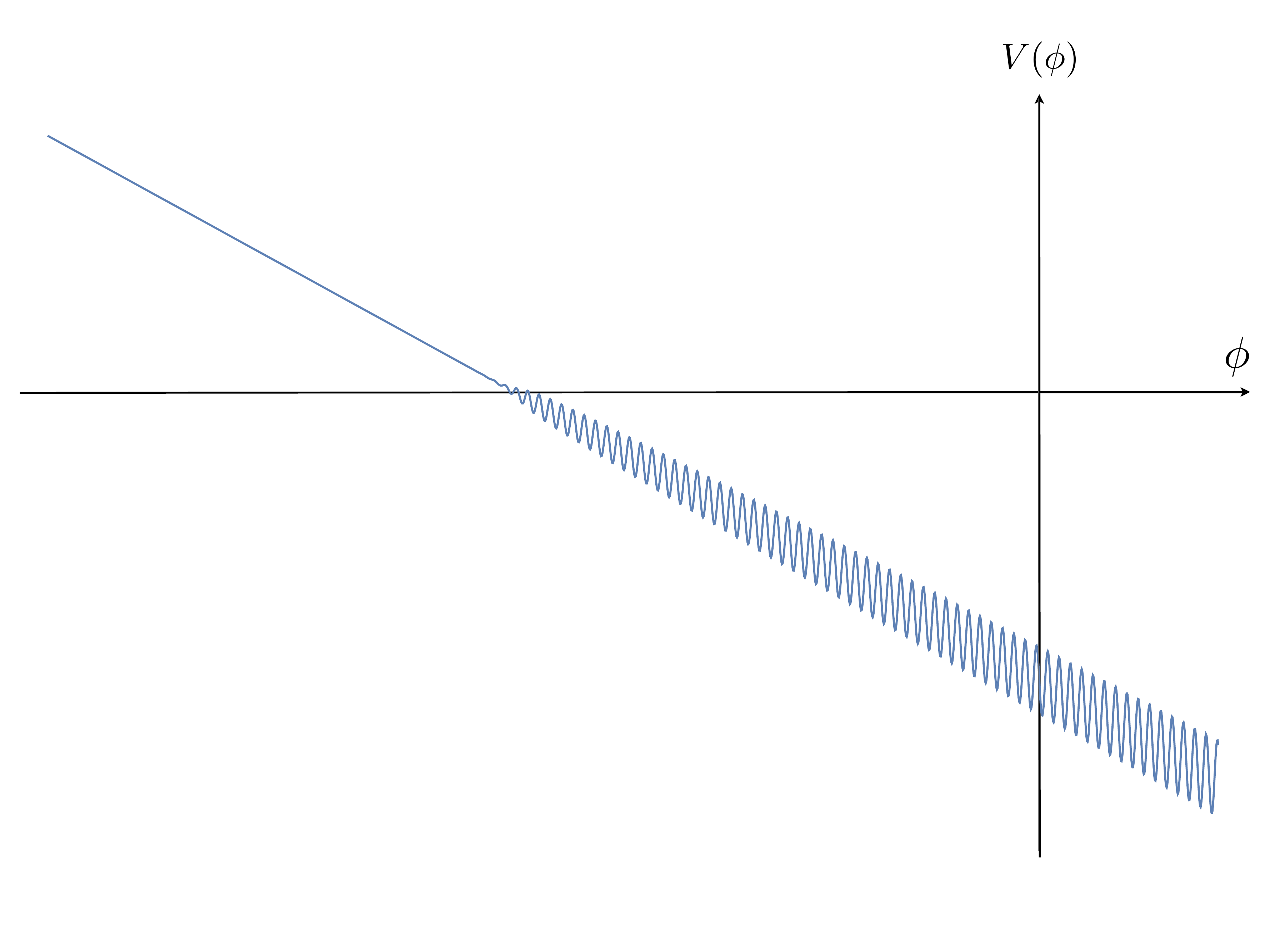}
\caption{Here is a characterization of the $\phi$'s potential in the region where the barriers begin to become important.  This is the one-dimensional slice in the field space after the Higgs is integrated out, effectively setting it to its minimum.  To the left, the Higgs vev is essentially zero, and is ${\cal O}(m_{\rm W})$ when the barriers become visible.  The density of barriers are greatly reduced for clarity.}
\label{potential-roll}
\end{center}
\end{figure}

We will now examine the dynamics of this model in the early universe.  We take an initial value for $\phi$ such that the  effective mass-squared of the Higgs, $m_h^2$, is positive.  During inflation, $\phi$ will slow-roll, thereby scanning the physical Higgs mass.  At some point in the $\phi$ potential, the quadratic term for the Higgs crosses zero and the Higgs develops a vacuum expectation value.  As the Higgs vev grows, the effective heights of the bumps, $\Lambda^4$, in the periodic potential grow.   When the bumps are large enough they become barriers which stop the rolling of $\phi$ shortly after $m_h^2$ crosses zero.  This sets the Higgs mass to be naturally much smaller than the cutoff (see Figure \ref{potential-roll}).  Since it is the axion which is responsible for the dynamical relaxation of the weak scale, we call it the relaxion.

The (rel)axion barrier height depends on the Higgs vev through its dependence on quark masses  \cite{Weinberg:1977ma}.  When the Higgs vev is near its standard model value, the potential barrier is approximately
\begin{equation}
\Lambda^4 \sim f_\pi^2  m_\pi^2
\end{equation}
times dimensionless ratios of quark masses.  Since $m_\pi^2$ changes linearly with the quark masses it is proportional to the Higgs vev.  Therefore $\Lambda^4$ grows linearly\footnote{We are ignoring the logarithmic contribution from the running of the QCD scale.  Also, when the Higgs mass squared is positive, and the vev is tiny due to QCD effects mixing with the Higgs, a better description of the pion is to say it is eaten by the weak gauge bosons.  Nevertheless, in that regime the coefficient of this potential still depends on the Higgs mass, but is negligible.} with the vev.

During inflation, the relaxion must roll over an $\mathcal{O}(1)$ fraction of its full field range, $\sim (M^2/g)$, to naturally cross the critical point for the Higgs where $m_h^2 = 0$.
Note that for the early universe dynamics, one can consider the potential to be just $g M^2 \phi$ or $g^2 \phi^2$ since the field value for $\phi \sim (M^2/g)$ makes these equivalent.
Our solution is insensitive to the initial condition for $\phi$ (as long as the Higgs starts with a positive mass-squared), because $\phi$ is slow-rolling due to Hubble friction.  This places the slow-roll constraints on $\phi$ that $g < \Hi$ and $g < (\Hi^2 \Mpl/M^2)$, where $\Hi$ is the Hubble scale during inflation and $\Mpl$ is the reduced Planck mass.  It will turn out that these constraints are trivially satisfied.  A requirement on inflation is that it lasts long enough for $\phi$ to scan the entire range.  During $N$ e-folds of inflation, $\phi$ changes by an amount $\Delta \phi \sim (\dot{\phi}/\Hi) N \sim (V'_\phi/\Hi^2) N \sim (g M^2/\Hi^2) N$.  Requiring that $\Delta \phi \gtrsim (M^2/g)$ gives the requirement on $N$
\begin{equation}
\label{eqn: efolds required}
N \gtrsim \frac{\Hi^2}{g^2}.
\end{equation}
There are three conditions on the Hubble scale of inflation.  First is that the vacuum energy during inflation is greater than the vacuum energy change along the $\phi$ potential, namely $M^4$, so
\begin{eqnarray}
\label{eqn: inflaton energy}
\Hi > {M^2 \over \Mpl} && ({\rm vacuum \; energy})
\end{eqnarray}
The second constraint is the requirement that  the Hubble scale during inflation is lower than the QCD scale (so the barriers form in the first place):
\begin{eqnarray}
\label{eqn: barriers form}
\Hi < \Lambda_{\rm QCD} &&({\rm barriers \; form}) 
\end{eqnarray}
where $\Lambda_{\rm QCD}$ is taken to be the scale where the instanton contributions to the axion potential are unsuppressed.  We expect numerically, $\Lambda_{\rm QCD}\sim \Lambda $.  
Finally, a condition could be placed on the Hubble scale by requiring that $\phi$'s evolution be dominated by classical rolling (and not quantum fluctuations -- similar to a constraint of $\delta\rho/\rho < 1$ in inflation) so that every inflated patch of the universe makes it to the electroweak vacua
\begin{eqnarray}
\label{eqn: quantum}
\Hi < \frac{V'_\phi}{\Hi^2} \rightarrow \Hi < (gM^2)^\frac{1}{3} && ({\rm classical \; beats \; quantum}) 
\end{eqnarray}
We'll see below that this constraint will be a bit stronger than the previous one, but in some cases it can be avoided.


The slow-rolling of $\phi$ stops when $\Lambda$ has risen to the point such that the slope of the barriers $\Lambda^4/f$ matches the slope of the potential, $g M^2$.  This occurs at
\begin{equation}
\label{eqn: barrier height}
g M^2 f \sim \Lambda^4.
\end{equation}
From the three conditions Eqns.~\eqref{eqn: inflaton energy}, \eqref{eqn: quantum}, and \eqref{eqn: barrier height}, we have a constraint on the cutoff $M$:
\begin{equation}
M < \left(\Lambda^{4} \Mpl^{3} \over f \right)^\frac{1}{6} \sim 10^7 \, \GeV \times  \left({10^9 \GeV \over f}\right)^\frac{1}{6}
\end{equation}
where we have scaled $f$ by its lower bound of $10^9 \GeV$ set by astrophysical constraints on the QCD axion (see for example \cite{Agashe:2014kda}).

\begin{figure}[htbp]
\begin{center}
\includegraphics[width=0.75 \textwidth]{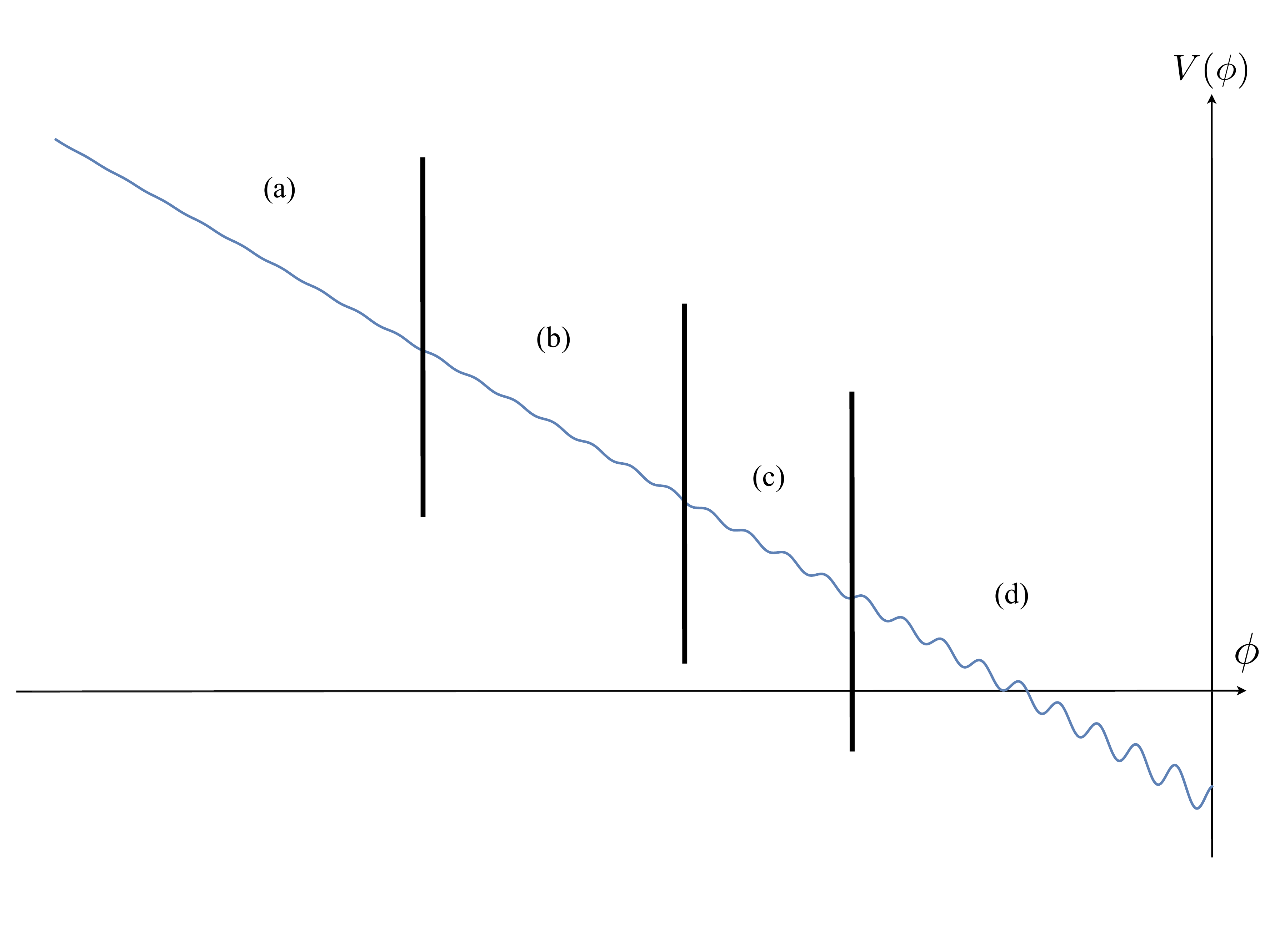}
\caption{A close up of the region of $\phi$'s potential as the barriers appear.  The evolution in these regions are (a) classical rolling dominated, (b) dominated by quantum fluctuations in the steps but classical rolling between steps, (c) classically stable, but quantum fluctuations/tunneling rates shorter than $N$ e-folds, and (d) classically stable, quantum transition rates longer than both $N$ e-folds and 10 Gyr.  Again, for clarity, the potential is not to scale.}
\label{potential-closeup}
\end{center}
\end{figure}

Note that in order to have a cutoff $M$ above the weak scale, $m_{\rm W}$, Eqn.~\eqref{eqn: barrier height} requires $gf\ll m_{\rm W}^2$.  This implies that the effective step size of the Higgs mass from one minimum to the next is much smaller than the weak scale.  So the barriers grow by a tiny fractional amount compared to $\Lambda_\text{QCD}$ per step.  Classically $\phi$ stops rolling as soon as the slope of its potential changes sign.  However since $gf \ll m_{\rm W}^2$, the slope of the first barrier after this point is exceedingly small, much smaller than $\Lambda^4/f$.  Therefore around this point, quantum fluctuations of $\phi$ will be relevant.  The field $\phi$ will be distributed over many periods $f$ (see Figure \ref{potential-closeup}), but in all of these the Higgs will have a weak-scale vev.  This quantum spreading is an oddity of the model.  As the universe inflates, different patches of the universe will have a range of $\phi$ field values and a range of Higgs vevs, but all around the weak scale.  In future work, we will show it is possible to build models which land the full initial patch in a single vacuum, thus removing this feature of our solution \cite{FutureGlory}.

Some of the resulting $\phi$ range is before the classical stopping point and is therefore classically unstable.
The rest is in  $\phi$ vacua with varying potential barrier heights.  
Far enough beyond the classical stopping point, $\phi$ reaches barriers where the slope pushing $\phi$ backwards towards a minimum is $\mathcal{O}(1)$ of the original slope $g M^2$.  By this point the quantum jumps of size $\Hi$ can no longer walk $\phi$ out of each minimum.  The lifetime of these vacua is much longer than the current age of the universe because tunneling rates are exponentially suppressed.  
In addition, if inflation lasts longer than $\sim 10$ Gyr (typical in our parameter space), this will easily guarantee that most patches populate the stable-enough vacua.  
Therefore, it is highly likely to end up in a patch of the universe which is at the weak scale and lives much longer than 10 Gyr.    
As a result of these multiple vacua, there will be domain walls after reheating in the full initial patch of the universe.  However these domain walls will be spaced by distances much larger than our current Hubble size because we have much more than 60 e-folds of inflation in any one vacuum, and are therefore unlikely to be observable.

We wish to avoid eternal inflation in our scenario because at least some part of the universe would end up with a Higgs vev above the weak scale.  The decay rates to such vacua are exponentially suppressed but with a long enough period of inflation, some fraction of the universe would end up there before reheating.  Although this might naively seem like a very small part of the universe, if we wish to avoid discussion of measures in eternal inflation we must avoid this possibility.  To do so, we can impose the constraint in Eq.~\eqref{eqn: quantum}, in which case the entire initial patch has the correct order electroweak scale at the end of inflation.

As noted above, even if we do not have eternal inflation in patches with large Higgs masses, we unfortunately cannot avoid ending up in a large range of vacua.  Since all of these vacua have weak scale Higgs vevs, we call this a solution to the hierarchy problem.  Of course, we have not solved the cosmological constant (CC) problem.  This set of final vacua will all have different cosmological constants.  If the solution to the CC problem is just tuning, then we must live in the one with the correct CC.  This is just the usual tuning required for the CC problem, and not an additional tuning.  Note that the other vacua with positive CC will eternally inflate (as is our universe presumably), but in any case have a  weak scale Higgs vev for a period that lasts much longer then 10 Gyr.

The model above is ruled out by the strong CP problem.  Since $\phi$ is the QCD axion its vev determines the $\theta$ parameter in QCD.  The relation Eq.~\eqref{eqn: barrier height} which determines where $\phi$ stops rolling predicts that the local minimum for $\phi$ is displaced from the minimum of the QCD part of the potential by ${\cal O}(f)$.  Therefore it generates $\theta \sim 1$.  We found two solutions to this problem:
\begin{enumerate}
\item{Potential barriers for $\phi$ arise from a new strong group, not QCD}
\item{The slope of the $\phi$ potential decreases dynamically after inflation}
\end{enumerate}
We discuss the latter solution below and the former in Section \ref{sec: non-qcd}.  Of course other solutions to the strong CP problem in this context would be interesting.

One way to decrease the slope after inflation is to tie it to the value of the inflaton $\sigma$.  We can add the term $\kappa \sigma^2 \phi^2$ to the potential.  One can check that our parameter space will remain technically natural, essentially because, like the relaxion, the classical value of the inflaton will be large compared to the cutoff.  There is now an additional slope, $\kappa \sigma^2 M^2/g$ which we take to be larger than $g M^2$.  Assuming $\sigma$ has a roughly constant value during most of inflation, we will describe this with a new effective coupling $\tilde{g}^2 = \kappa \sigma^2$ which is constant during most of inflation.  The inflation field drops to zero after inflation, removing this new contribution to the potential and leaving the original slope $\sim g M^2$.  In order to solve the strong CP problem as well, we need the slope of the potential to drop by a factor of $\theta \lesssim 10^{-10}$ after inflation so that the axion is only displaced by this amount from its (local) minimum.  Thus we require $g M^2 \sim \theta \tilde{g}^2 (M^2/g)$, or $g \sim \tilde{g} \sqrt{\theta}$.
This has the added benefit that once the slope drops, every $\phi$ vacuum that any patch of the universe sits in now becomes very long lived because the effective barriers rose by $\gtrsim 10^{10}$.  It is easy to show that quantum corrections from this term (assuming $\sigma>\Mpl$ do not contribute significantly to the $\phi$ potential.

The condition on the number of e-folds of inflation is now
\begin{equation}
\label{eqn: efolds required2}
N \gtrsim \theta\frac{\Hi^2}{g^2}.
\end{equation}
The conditions Eqns.~\eqref{eqn: inflaton energy}, \eqref{eqn: barriers form}, \eqref{eqn: quantum}, and \eqref{eqn: barrier height} become respectively
\begin{eqnarray}
\label{eqn: inflaton energy2}
\Hi > \frac{M^2}{\Mpl \sqrt{\theta}} && ({\rm vacuum \; energy}) 
\\
\nonumber
\\
\label{eqn: barriers form2}
\Hi < \Lambda_{\rm QCD} &&({\rm barriers \; form}) 
\\
\label{eqn: quantum2}
\Hi < \left( \frac{g M^2}{\theta} \right)^\frac{1}{3} && ({\rm classical \; beats \; quantum})
\\
\nonumber
\\
\label{eqn: barrier height2}
g M^2 f \sim \Lambda^4 \theta && ({\rm barrier \; heights}) 
\end{eqnarray}
Note that because of the dropping slope, the vacuum energy Eq.~\eqref{eqn: inflaton energy2} is greater than the fourth power of the cutoff, $M^4$, by a power of $\theta^{-1}$.  This is not a problem for the effective theory, but may be a concern for the UV completion.

The constraints above give a bound on the cutoff of
\begin{equation}
\label{eqn: QCD cutoff}
M < {\left(\Lambda^{4} \Mpl^{3} \over f \right)}^{\frac{1}{6}} \theta^\frac{1}{4} \sim 30 \TeV \times  {\left({10^9 \, \GeV \over f}\right)}^\frac{1}{6} {\left( \frac{\theta}{10^{-10}} \right)}^\frac{1}{4}
\end{equation}
This model now satisfies all constraints and has only the QCD axion and inflaton added to the Standard Model below the cutoff.  Thus the minimal model has no hierarchy problem, no strong CP problem, and has a natural candidate for dark matter.  Due to the constraints, its full parameter space can be probed in a number of ways in future experiments.

Our mechanism for solving strong CP, dropping the slope, potentially allows us to loosen the constraint from requiring classical rolling to dominate, Eq.~\eqref{eqn: quantum2}.  If when the slope drops, the sign of the underlying slope is the opposite sign,  the small fraction of patches that have not reached the barriers will eventually find themselves with a large negative cosmological constant and should suffer a collapse and not eternal inflation (assuming one of the 'typical' patches has our measured cosmological constant).  Potentially we can ignore the classical rolling constraint, and allow a higher cutoff -- using Eq.~\eqref{eqn: inflaton energy2} and Eq.~\eqref{eqn: barriers form2}, we derive
\begin{equation}
\label{eqn: new QCD cutoff}
M < {\left(\Lambda \Mpl \right)}^{\frac{1}{2}} \theta^\frac{1}{4} \sim 1000 \TeV \times {\left( \frac{\theta}{10^{-10}} \right)}^\frac{1}{4}
\end{equation}
One concern about loosening this constraint is that there are a small fraction of patches that naively fluctuate well beyond $\phi \sim M^2/g$.  These patches, however, are not a concern if, for example, $\phi$ is periodic with period $\sim M^2/g$.

A final consistency check is making sure that reheating at the end of inflation does not destabilize $\phi$ and take us out of a good minimum (one in which the electroweak scale is the correct size).  If the standard model fields reheat to temperatures below the temperature where appreciable barriers form (roughly 3 GeV assuming $\theta= 10^{-10}$ -- see, for example, \cite{Wantz:2009it}), then $\phi$ remains in its original vacuum, though it can be displaced from its minimum.  If the reheat temperature is above this scale, the barriers effectively disappear and the relaxion can begin to roll.  We estimate the distance $\phi$ rolls (and one can show it slow-rolls as long as $\Delta\phi<\Mpl$) as:
\begin{equation}
\frac{\Delta\phi}{f} \sim \frac{\dot{\phi}}{H_b f} \sim \frac{V' }{H_b^2 f} \sim \theta \frac{\Lambda^4}{T_b^4} \frac{\Mpl^2}{f^2}
\end{equation}
Where $T_b$ is the scale where barriers begin to form and $H_b$ is the Hubble scale at this temperature.  Taking $T_b \sim 3 \GeV$ and $\theta= 10^{-10}$, we find $\phi$ moves less than one period if $f>10^{10}\GeV$.  However, even for $f = 10^9 \GeV$ (roughly the lower bound on the QCD axion coupling), while the relaxion rolls through multiple periods, it changes the Higgs squared mass by less than 1 eV$^2$ for any cutoff above 1 TeV.  And because the relaxion still easily satisfies the slow roll condition, it stops rolling once the barriers appear.


\section{Non-QCD model}
\label{sec: non-qcd}


Our solution to the hierarchy problem only requires the Higgs vev to produce barriers which stops $\phi$ from rolling.  If the barriers are produced by something other than QCD, we can avoid the impact on the strong CP problem (as it can, for example, be solved by the standard axion), and the barrier heights can be larger than the QCD scale.  As we see in the model below, both of these allow for a larger upper bound on the cutoff, though we require a coincidence of scales due to current experimental constraints (similar to the $\mu$-problem in the minimal supersymmetric standard model \cite{Dimopoulos:1981zb}).

The dynamics of this model are similar to the previous one -- $\phi$ rolls until the Higgs vev is large enough to produce barriers to stop $\phi$.  The $\phi$ Lagrangian is the same as in the first model, except that it couples to the ${\tilde G}'^{\mu\nu}G'_{\mu\nu}$ of a new strong group (not QCD), which we take to be $SU(\bf{3})$.
 The Higgs couples to new fermions which are charged under both the new strong group, and the electroweak group.  It's vev contributes to their masses and raises the barriers when turned on.
 The upper bound on the cutoff is much larger than the model in Section \ref{sec: QCD}, mostly due to the avoidance of the strong CP contributions.
 The new fermions are required to be at the weak scale, and thus are collider accessible and impact Higgs and electroweak precision physics.
	
The new fermions are labelled suggestively as ($L,N$) and their conjugates ($L^c,N^c$). The fields $L$ and $N$ carry the same standard model charges as the lepton doublet and right handed neutrino respectively, and are in the fundamental representation of the new strong group, and $L^c$ and $N^c$ are in the conjugate representations.  They have Dirac masses and Yukawa couplings with the Higgs as follows:
\begin{equation}
{\cal L} \supset  m_L L L^c + m_N N N^c  + y h L N^c + \tilde{y} h^\dagger L^c N
\label{Eq:stronghiggs}
\end{equation}	
Collider and other constraints require $m_L$ to be greater than the weak scale, but no such constraint exists on $m_N$, and the barriers in the $\phi$ potential vanish as the lightest fermion mass goes to zero.  Thus the key is that a Higgs vev can significantly increase the mass of the lightest fermion at tree-level.  A naive dimensional analysis estimate of the barrier coefficient (in front of the periodic potential) is $\Lambda^4 \simeq 4\pi f_{\pi'}^3 m_N$, where $f_{\pi'}$ is the chiral symmetry breaking scale of the new strong group and we have assumed $m_L \gg f_{\pi'} \gg m_N$.

In this limit, the Higgs vev gives a contribution to the lightest fermion mass of size $y \tilde{y} \vev^2 / m_L$.  Technical naturalness requires $N$'s Dirac mass to be at least the larger of $\sim (y \tilde{y}/16\pi^2) m_L \log{M/m_L}$ and $\sim y \tilde{y} f_{\pi'}^2/m_L$, and thus the Higgs vev only has a significant impact if 
\begin{equation}\label{eq:mL-bound}
f_{\pi'}<\vev {\rm \;\;\; and \;\;\;} m_L < {4\pi \vev\over \sqrt{\log{M/m_L}}}.
\end{equation}
In addition, for the Higgs to have an effect, the lightest fermion, of mass $\sim m_N$, should be lighter than the confinement scale -- otherwise the axion potential will be saturated.  The additional constraint is
\begin{equation}\label{eqn: mN-bound}
4\pi f_{\pi'} > {y \tilde{y} \vev^2 \over m_L}.
\end{equation}

There should be a lower limit on $m_L$ around the weak scale from collider production of $L,L^c$.  In the part of parameter space with the largest allowed $f_{\pi'}$ (and largest allowed cutoff), the bound should be weaker than that on chargino/neutralino production \cite{colliders} as only the baryon-like states should leave significant missing transverse energy, while the meson states decay promptly via mixing with the Higgs.  
Another constraint on the Yukawa couplings is from Higgs physics, namely decays of the Higgs to the composite $N$ states.  For example, if $y,\tilde{y}\lesssim 0.1$, and $m_L> 250 \GeV$, the branching ratio to the new mesons is less than $10\%$.  In addition, there are precision electroweak constraints, which are more important than the Higgs constraints only if $m_L$ is small.
Finally, there may be interesting cosmological constraints (or signals) on higher-dimensional operators from the long-lived or stable baryons in this sector.  We leave all of these studies for future work.

Thus, the dynamics are exactly those of the model in Section \ref{sec: QCD}, where $\phi$ rolls, turns on the Higgs vev, and is stopped by barriers determined by the vev.  The same constraints in Equations \eqref{eqn: inflaton energy}, \eqref{eqn: quantum}, and \eqref{eqn: barrier height} apply (in which case, \eqref{eqn: barriers form} is already satisfied), with $\Lambda^4 \rightarrow 4\pi f_{\pi'}^3 m_N \sim 4\pi f_{\pi'}^3 y \tilde{y} \vev^2 / m_L$.  One additional difference is that the $\phi$ field is no longer the QCD axion, and so the bounds on its couplings are much weaker.  Assuming $f$ is at least as large as the cutoff, we can parameterize the bound on $M$ as
\begin{eqnarray}
M &< &(\Lambda^4 \Mpl^3)^{1\over 7}  \left({M \over f}\right)^{1\over 7} \\ \nonumber \\ \nonumber
 &<& 3\times 10^8 \GeV \left({f_{\pi'}\over 30 {\rm \; GeV}}\right)^{3\over 7} \left({y \tilde{y} \over 10^{-2}}\right)^{1\over 7} \left({300 {\rm \; GeV}\over m_L} \right)^{1\over 7} \left({M \over f}\right)^{1\over 7} \\ \nonumber \\ \nonumber
 &<& 2 \times 10^8 \GeV \left({f_{\pi'}\over 30 {\rm \; GeV}}\right)^{4\over 7} \left({M \over f}\right)^{1\over 7}
\end{eqnarray}
where in the last line, we used Eq.~\eqref{eqn: mN-bound}.  In the standard model, a cutoff that saturates this bound would require a tuning of one part in $10^{12}$.  Here, we have achieved this hierarchy dynamically.

Again, a final constraint comes if reheating occurs above the strong coupling scale.  In that case, the relaxion begins to roll, and unlike the QCD case, requiring slow roll is a non-trivial constraint.  Slow roll requires
\begin{equation}
\label{eqn: reheat slow roll}
\epsilon = {1\over 2} \left(\frac{V'}{H_b^2 \Mpl}\right)^2 \simeq {1\over 2} \left(\frac{\Lambda^4}{T_b^4} \frac{\Mpl}{f}\right)^2 \ll 1
\end{equation}
where again, $T_b$ is the temperature at which the effective barriers appear.  We estimate this to be $T_b^4 \sim 16 \pi^2 f_{\pi'}^4$ (to match the corresponding temperature in QCD).  Using this and our formulas for $\Lambda^4$ and the lightest fermion, $m_{N}$, we arrive at a lower bound on $f$:
\begin{equation}
f > 10^{-2}\Mpl \left(\frac{y\tilde{y}}{10^{-2}}\right) \left(\frac{30\GeV}{f_{\pi'}}\right) \left(\frac{300\GeV}{m_L}\right) 
\end{equation}
The requirement that $\phi$ not roll outside the range of vacua with weak scale Higgs vev is a weaker requirement (for any cutoff above TeV).

Note, from Equation (\ref{eq:mL-bound}) this model ceases to work properly if either $m_L$ or $f_{\pi'}$ gets much above a few hundred GeV.  Thus, in this model, we see that a natural solution to the hierarchy problem requires the existence of new weak scale electroweak particles charged under a new gauge group which confines below the weak scale.  However, these particles need not be charged under QCD, making them harder to detect at hadron colliders.  In addition, while precision Higgs and electroweak observables depend strongly on the Yukawa couplings, $M$ depends only weakly on them, and thus constraints can be easily evaded without significant effect on the parameter space.


\section{Example Inflation Sector}
\label{sec:inflation}


We need many e-folds of inflation in order to have enough time for the scanning of the Higgs mass.  We find it preferable to avoid eternal inflation because then a multiverse is produced which will ultimately populate all our vacua.  Even without eternal inflation though, most inflation models can easily produce many e-folds.  For example, even single field inflation with a $m^2 \sigma^2$ potential (where $\sigma$ is the inflaton) will produce enough e-folds with the required low Hubble scale when $m \sim 10^{-27}$ GeV.  However, it would have to be followed by a second stage of inflation to achieve the observed $\delta \rho / \rho$ and a large enough reheat temperature.  It is not surprising that single-field inflation can achieve the required number of e-folds since the constraints on our models are very similar to those on inflation.

In this section we give a simple hybrid inflation model as a proof of principle that achieves all our requirements on the inflation model and gives the observed $\delta \rho / \rho$.
As is a generic issue with many low-scale inflation models however, this inflation sector is not natural.
We will demonstrate a model for the QCD axion solution.  The same model works for the non-QCD axion solution, and has fewer constraints.
In the future we will present a new type of inflation sector based on our mechanism which is natural and satisfies all the constraints necessary for our solution to the hierarchy problem \cite{FutureGlory}.
It would be interesting to find other natural models of inflation that also satisfy our constraints.

We consider a hybrid inflation sector \cite{Linde:1993cn}, with the following relevant terms in the scalar potential:
\begin{equation}
V \ni m^2 \sigma^2 + c \sigma^2 \chi^2 - m_\chi^2 \chi^2 + \lambda \chi^4
\end{equation}
where $\sigma$ is the inflaton and $\chi$ is the waterfall field.
We must satisfy the constraints on the inflation model in Eqns.~\eqref{eqn: efolds required2} and \eqref{eqn: inflaton energy2}.  We will take an initial phase of inflation with super-Planckian field excursion for $\sigma$ which is followed by a normal hybrid inflation phase driven by the energy in $\chi$.  Further, we require $\delta \rho / \rho < 1$ at the beginning of inflation in order to avoid eternal inflation.  And observations require $\delta \rho / \rho \approx 10^{-5}$ by the end of inflation.
Putting all these constraints together leaves an open parameter space.    One set of parameters which work for the QCD axion model are $M \sim 10^4 \, \GeV$, $f \sim 10^9 \, \GeV$, $\Lambda \sim 10^{-1} \, \GeV$, $g \sim 10^{-31} \, \GeV$, $\theta \sim 10^{-10}$, $\Hi \sim 10^{-5} \, \GeV$, final Hubble scale $H_f \sim 10^{-12} \, \GeV$ and $\lambda \sim 10^{-1}$.  Instead of attempting to characterize the entire parameter space, we simply present this one point which works since our goal is just to illustrate that it is possible to find an inflation sector for our model.
One could even attempt to make this model natural by supersymmetrizing it, but this model is just meant to demonstrate that the requirements on inflation are potentially satisfiable and for example it does not even predict an allowed scalar tilt\footnote{We thank Renata Kallosh for pointing this out.}.  Thus, significant progress is still to be made with a viable inflation sector.


\section{Observables}


Central to our class of solutions is a new, light, very-weakly coupled boson.  The most promising ways to detect this field are through low-energy, high-precision experiments.  This is in stark contrast to conventional solutions to the hierarchy problem which require new physics at the weak scale and hence are (at least potentially) observable in colliders.
 A comprehensive discussion of the experimental program necessary to discover this mechanism is beyond the scope of this work - we will instead highlight experimental strategies that seem promising.  
While it may be challenging to ultimately confirm our mechanism, it is an open goal which will hopefully motivate new types of searches.




Our class of solutions generically predicts axion-like dark matter.  The simplest model predicts the QCD axion as a dark matter candidate.  Excitingly, a new area of direct detection experiments focused on light bosons is now emerging \cite{Graham:2011qk, Graham:2013gfa, Budker:2013hfa, Sikivie:1983ip, Asztalos:2009yp, Graham:2014sha, Chaudhuri:2014dla, Graham:2015rva, Arvanitaki:2014dfa, Arvanitaki:2014faa, Derevianko:2013oaa, An:2014twa, Izaguirre:2014cza, An:2013yua, Pospelov:2012mt, Arias:2014ela, Arias:2012az, Horns:2012jf, Stadnik:2013raa, Stadnik:2014xja}.  These new experiments may, for example, open up the entire QCD axion range to exploration.  In the parts of parameter space where the axion-like particle's lifetime is at or below the age of the universe, there will already be constraints or potential cosmological signals (see for example \cite{Cadamuro:2011fd}).
It is interesting that light field (axion-like) dark matter candidates in our theories replace the heavy particle (WIMP-like) candidates of conventional solutions to the hierarchy problem.

While our theories can have axion dark matter, the specific prediction for the axion abundance and mass-coupling relation may be altered.  Because the axion potential now has an overall slope, it can acquire an initial velocity in the early universe after reheating set by the slow-roll condition.  This would change the calculation of the final axion dark matter abundance and is thus important to work out.  We leave this for future work, but note that this could predict QCD axion dark matter in a region of parameter space different from where the standard axion model does.  In addition, in the non-QCD case, the field $\phi$ may be stopped right when the barriers first appear and therefore the mass of the axion particle may be naturally tuned to be small.  This small mass improves the observability of the axion dark matter \cite{Graham:2013gfa, Budker:2013hfa}.  Interestingly if the axion is observed, its mass and couplings can be measured and would not satisfy the usual relation with the confinement scale $\Lambda$ (potentially measurable in colliders).  Observation of such dark matter would be a tantalizing hint of our mechanism. 

In the QCD case there is a preference for large $\theta$ from Eqn.~\eqref{eqn: QCD cutoff}.  While this is a relatively weak preference because of the $1/4$ power, it does favor
a static nucleon EDM that may be observable\footnote{This is in addition to the oscillating EDM induced by the axion in this scenario \cite{Graham:2011qk}}.  Upcoming nucleon EDM experiments are predicted to improve on the current bounds by several orders of magnitude, potentially providing further hints of this scenario (see for example \cite{Anastassopoulos:2015ura}).  In the non-QCD case there can be a large theta in the new strong group.  A two-loop diagram may then give EDMs for nucleons or electrons which could be detectable and may even give a constraint on parameter space.

Our models appear to generically require low scale inflation (unless we find a new dissipation mechanism besides Hubble friction during inflation).  This prediction can be falsified by observation of gravitational waves from inflation, but it cannot be directly observed.

The models presented in this paper either have a low cutoff (in the QCD case) or new physics at the weak scale (in the non-QCD case).  Either case is then potentially observable at the LHC or future colliders.  The non-QCD case has new fermions at the weak scale charged under a new strong group with a confinement scale below the weak scale.  This scenario should have rich phenomenology, for example the lightest states are composite singlet scalars that can mix with the Higgs.  For compositeness scales much smaller than the weak scale, the phenomenology may be similar to \cite{Juknevich:2009gg, Burdman:2014zta, Craig:2015pha}.   Both direct searches for new fermions with electroweak quantum numbers, as well as more refined measurements of Higgs branching ratios could probe the parameter space of this model, though the latter can be suppressed with small Yukawa couplings without significantly impacting our bounds.  Because the non-QCD model fails to be effective without electroweak fermions with masses in the hundreds of GeV, the whole parameter space could conceivably be covered by the LHC and/or a future linear collider.  Further studies of optimal strategies are warranted.  Observation of this new weak scale physics 
could provide the first evidence of such a mechanism.

Verification of a critical piece of this class of theories could come by observing the direct coupling of the new light field to the Higgs.  While this is unlikely to happen in colliders, there may be significant opportunities in new low-energy experiments.
As a component of dark matter, oscillations of the new light field cause oscillations of the Higgs vev.  This causes all scales connected to the Higgs, for example the electron mass, to oscillate in time with frequency equal to the axion mass.  Additionally the new light field couples to matter through its mixing with the Higgs and so mediates a new force.  It may be possible to design new high-precision experiments to search for these phenomena \cite{newexperiments}.  Such searches will be quite challenging.  However, if axion-like dark matter is discovered first, and thus its mass is measured, that mass can be targeted greatly enhancing the sensitivity of resonance searches  \cite{newexperiments}.


\section{Discussion}

We have found a new class of solutions to the hierarchy problem.  The two models in this paper are examples of a broader class of theories in which dynamical evolution in the universe drives the weak scale to its small value.  We find that in order to realize a model of this type it is necessary to satisfy the following conditions.
\begin{enumerate}
\item
Dissipation - Dynamical evolution of a field requires energy transfer which must be dissipated in order to allow the field to stop, and hence stop the scan of the Higgs mass.  This also allows the model to be insensitive to initial conditions.  In our models, dissipation is accomplished by gravity via hubble friction during inflation.

\item
Self-similarity - Cutoff-dependent quantum corrections will choose an arbitrary point in the scanning field's range at which the Higgs mass is cancelled.  The scanning field must therefore have a self-similar potential across its entire field range so that the Higgs can stop its evolution at any arbitrary point.  In our models the periodic axion potential provides this self-similarity.

\item
Higgs back-reaction - The Higgs vev must back-react on the scanning field, stopping the evolution at the appropriate value.  In our models this is accomplished in a technically natural way by coupling the Higgs to fermions which affect the scanning field's potential.

\item
Long time period - There must be a sufficiently long time period during the early universe for the Higgs mass to be scanned across the entire range from the cutoff to zero.

\end{enumerate}
It would be valuable to find other models in this class \cite{FutureGlory}.
In a sense, this type of theory gives a specific realization of the hope of applying ``self-organized criticality" to the weak scale \cite{Giudice:2008bi}.

More can be learned about this class of theories by finding ultraviolet completions.  UV completions may impose additional constraints on these models but may also reveal new realizations of this mechanism (e.g. as realized in string theory with axion monodromy or higher dimensional effective field theory \cite{Silverstein:2008sg, McAllister:2008hb, Kaloper:2008fb, Kaloper:2011jz, delaFuente:2014aca}).

While we have used this mechanism for the hierarchy problem, it is possible that it could be applied to other naturalness problems.  For example, instead of the Higgs boson, it could be used to make other scalar fields light (the inflaton, curvaton, chameleons, etc.).   Of course the biggest naturalness problem is the fine-tuning of the cosmological constant.  Perhaps a variant of the mechanism could lead the way to a new solution.

\section*{Acknowledgements}
We would like to thank Nathanial Craig, Savas Dimopoulos, Roni Harnik, Shamit Kachru, Renata Kallosh, Nemanja Kaloper, Jeremy Mardon, Aaron Pierce, Prashant Saraswat, Eva Silverstein, Raman Sundrum, Tim Tait,  Scott Thomas, and Jerry Zucker
for useful discussions. PWG acknowledges the support of NSF grant PHY-1316706, DOE Early Career Award DE-SC0012012, and the Terman Fellowship.  DEK acknowledges the support of NSF grant PHY-1214000.  SR acknowledges the support of NSF grant PHY-1417295.


\begin{thebibliography}{10}
\expandafter\ifx\csname url\endcsname\relax
  \def\url#1{{\tt #1}}\fi
\expandafter\ifx\csname urlprefix\endcsname\relax\def\urlprefix{URL }\fi

\bibitem{Wilson}
K.~Wilson, {\it unpublished}

\bibitem{Dimopoulos:1981zb} 
  S.~Dimopoulos and H.~Georgi,
  Nucl.\ Phys.\ B {\bf 193}, 150 (1981).

\bibitem{Randall:1999ee} 
  L.~Randall and R.~Sundrum,
  Phys.\ Rev.\ Lett.\  {\bf 83}, 3370 (1999)
  [hep-ph/9905221].


\bibitem{ArkaniHamed:1998rs} 
  N.~Arkani-Hamed, S.~Dimopoulos and G.~R.~Dvali,
  Phys.\ Lett.\ B {\bf 429}, 263 (1998)
  [hep-ph/9803315].


\bibitem{Susskind:1978ms} 
  L.~Susskind,
  Phys.\ Rev.\ D {\bf 20}, 2619 (1979).

\bibitem{Weinberg:1979bn} 
  S.~Weinberg,
  Phys.\ Rev.\ D {\bf 19}, 1277 (1979).
 
  
\bibitem{Abbott:1984qf} 
  L.~F.~Abbott,
  Phys.\ Lett.\ B {\bf 150}, 427 (1985).

\bibitem{'tHooft:1980xb} 
  G.~'t Hooft in 
  ``Recent Developments in Gauge Theories. Proceedings, Nato Advanced Study Institute, Cargese, France, August 26 - September 8, 1979,''
  G.~'t Hooft, et al, ed., 
  NATO Sci.\ Ser.\ B {\bf 59}, pp.1 (1980).


\bibitem{Silverstein:2008sg} 
  E.~Silverstein and A.~Westphal,
  Phys.\ Rev.\ D {\bf 78}, 106003 (2008)
  [arXiv:0803.3085 [hep-th]].
  
\bibitem{McAllister:2008hb} 
  L.~McAllister, E.~Silverstein and A.~Westphal,
  Phys.\ Rev.\ D {\bf 82}, 046003 (2010)
  [arXiv:0808.0706 [hep-th]].
  
\bibitem{Kaloper:2008fb} 
  N.~Kaloper and L.~Sorbo,
  Phys.\ Rev.\ Lett.\  {\bf 102}, 121301 (2009)
  [arXiv:0811.1989 [hep-th]].
  
\bibitem{Kaloper:2011jz} 
  N.~Kaloper, A.~Lawrence and L.~Sorbo,
  JCAP {\bf 1103}, 023 (2011)
  [arXiv:1101.0026 [hep-th]].
  
\bibitem{delaFuente:2014aca} 
  A.~de la Fuente, P.~Saraswat and R.~Sundrum,
  Phys.\ Rev.\ Lett.\  {\bf 114}, no. 15, 151303 (2015)
  [arXiv:1412.3457 [hep-th]].



\bibitem{Peccei:1977hh} 
  R.~D.~Peccei and H.~R.~Quinn,
  Phys.\ Rev.\ Lett.\  {\bf 38}, 1440 (1977).

\bibitem{Weinberg:1977ma} 
  S.~Weinberg,
  Phys.\ Rev.\ Lett.\  {\bf 40}, 223 (1978).

\bibitem{Wilczek:1977pj} 
  F.~Wilczek,
  Phys.\ Rev.\ Lett.\  {\bf 40}, 279 (1978).
  
 
  
\bibitem{Agashe:2014kda} 
  K.~A.~Olive {\it et al.}  [Particle Data Group Collaboration],
  Chin.\ Phys.\ C {\bf 38}, 090001 (2014).
  
\bibitem{FutureGlory} 
 P.~W.~Graham,  D.~Kaplan  and S.~ Rajendran, to appear
 

\bibitem{Wantz:2009it} 
  O.~Wantz and E.~P.~S.~Shellard,
  Phys.\ Rev.\ D {\bf 82}, 123508 (2010)
  [arXiv:0910.1066 [astro-ph.CO]].
  
  
\bibitem{colliders}
See for example\\
  G.~Aad {\it et al.}  [ATLAS Collaboration],
  JHEP {\bf 1404}, 169 (2014)
  [arXiv:1402.7029 [hep-ex]].
  and \\
  V.~Khachatryan {\it et al.}  [CMS Collaboration],
  Eur.\ Phys.\ J.\ C {\bf 74}, no. 9, 3036 (2014)
  [arXiv:1405.7570 [hep-ex]].
  
\bibitem{Linde:1993cn} 
  A.~D.~Linde,
  Phys.\ Rev.\ D {\bf 49}, 748 (1994)
  [astro-ph/9307002].








  
  
\bibitem{Graham:2011qk} 
  P.~W.~Graham and S.~Rajendran,
  Phys.\ Rev.\ D {\bf 84}, 055013 (2011)
  [arXiv:1101.2691 [hep-ph]].
  
\bibitem{Graham:2013gfa} 
  P.~W.~Graham and S.~Rajendran,
  Phys.\ Rev.\ D {\bf 88}, 035023 (2013)
  [arXiv:1306.6088 [hep-ph]].
  
\bibitem{Budker:2013hfa} 
  D.~Budker, P.~W.~Graham, M.~Ledbetter, S.~Rajendran and A.~Sushkov,
  Phys.\ Rev.\ X {\bf 4}, no. 2, 021030 (2014)
  [arXiv:1306.6089 [hep-ph]].
  
\bibitem{Sikivie:1983ip} 
  P.~Sikivie [ADMX Collaboration],
  Phys.\ Rev.\ Lett.\  {\bf 51}, 1415 (1983)
  [Phys.\ Rev.\ Lett.\  {\bf 52}, 695 (1984)].
  
\bibitem{Asztalos:2009yp} 
  S.~J.~Asztalos {\it et al.}  [ADMX Collaboration],
  Phys.\ Rev.\ Lett.\  {\bf 104}, 041301 (2010)
  [arXiv:0910.5914 [astro-ph.CO]].
  
  
  
\bibitem{Graham:2014sha} 
  P.~W.~Graham, J.~Mardon, S.~Rajendran and Y.~Zhao,
  Phys.\ Rev.\ D {\bf 90}, no. 7, 075017 (2014)
  [arXiv:1407.4806 [hep-ph]].
  
\bibitem{Chaudhuri:2014dla} 
  S.~Chaudhuri, P.~W.~Graham, K.~Irwin, J.~Mardon, S.~Rajendran and Y.~Zhao,
  arXiv:1411.7382 [hep-ph].
  
\bibitem{Graham:2015rva} 
  P.~W.~Graham, J.~Mardon and S.~Rajendran,
  arXiv:1504.02102 [hep-ph].
  
\bibitem{Arvanitaki:2014dfa} 
  A.~Arvanitaki and A.~A.~Geraci,
  Phys.\ Rev.\ Lett.\  {\bf 113}, no. 16, 161801 (2014)
  [arXiv:1403.1290 [hep-ph]].
  
  
\bibitem{Arvanitaki:2014faa} 
  A.~Arvanitaki, J.~Huang and K.~Van Tilburg,
  Phys.\ Rev.\ D {\bf 91}, no. 1, 015015 (2015)
  [arXiv:1405.2925 [hep-ph]].
  
\bibitem{Derevianko:2013oaa} 
  A.~Derevianko and M.~Pospelov,
  Nature Phys.\  {\bf 10}, 933 (2014)
  [arXiv:1311.1244 [physics.atom-ph]].
  
\bibitem{An:2014twa} 
  H.~An, M.~Pospelov, J.~Pradler and A.~Ritz,
  arXiv:1412.8378 [hep-ph].
  
\bibitem{Izaguirre:2014cza} 
  E.~Izaguirre, G.~Krnjaic and M.~Pospelov,
  arXiv:1405.4864 [hep-ph].
  
\bibitem{An:2013yua} 
  H.~An, M.~Pospelov and J.~Pradler,
  Phys.\ Rev.\ Lett.\  {\bf 111}, 041302 (2013)
  [arXiv:1304.3461 [hep-ph]].
  
\bibitem{Pospelov:2012mt} 
  M.~Pospelov, S.~Pustelny, M.~P.~Ledbetter, D.~F.~Jackson Kimball, W.~Gawlik and D.~Budker,
  Phys.\ Rev.\ Lett.\  {\bf 110}, no. 2, 021803 (2013)
  [arXiv:1205.6260 [hep-ph]].

\bibitem{Arias:2014ela} 
  P.~Arias, A.~Arza, B.~Dšbrich, J.~Gamboa and F.~Mendez,
  arXiv:1411.4986 [hep-ph].
  
\bibitem{Arias:2012az} 
  P.~Arias, D.~Cadamuro, M.~Goodsell, J.~Jaeckel, J.~Redondo and A.~Ringwald,
  JCAP {\bf 1206}, 013 (2012)
  [arXiv:1201.5902 [hep-ph]].
  
\bibitem{Horns:2012jf} 
  D.~Horns, J.~Jaeckel, A.~Lindner, A.~Lobanov, J.~Redondo and A.~Ringwald,
  JCAP {\bf 1304}, 016 (2013)
  [arXiv:1212.2970].


\bibitem{Cadamuro:2011fd} 
  D.~Cadamuro and J.~Redondo,
  JCAP {\bf 1202}, 032 (2012)
  [arXiv:1110.2895 [hep-ph]].
  
\bibitem{Stadnik:2013raa} 
  Y.~V.~Stadnik and V.~V.~Flambaum,
  Phys.\ Rev.\ D {\bf 89}, no. 4, 043522 (2014)
  [arXiv:1312.6667 [hep-ph]].
  
\bibitem{Stadnik:2014xja} 
  Y.~V.~Stadnik and V.~V.~Flambaum,
  Eur.\ Phys.\ J.\ C {\bf 75}, no. 3, 110 (2015)
  [arXiv:1408.2184 [hep-ph]].


\bibitem{Anastassopoulos:2015ura} 
  V.~Anastassopoulos, S.~Andrianov, R.~Baartman, M.~Bai, S.~Baessler, J.~Benante, M.~Berz and M.~Blaskiewicz {\it et al.},
  arXiv:1502.04317 [physics.acc-ph].


\bibitem{Juknevich:2009gg} 
  J.~E.~Juknevich,
  JHEP {\bf 1008}, 121 (2010)
  [arXiv:0911.5616 [hep-ph]].

\bibitem{Burdman:2014zta} 
  G.~Burdman, Z.~Chacko, R.~Harnik, L.~de Lima and C.~B.~Verhaaren,
  Phys.\ Rev.\ D {\bf 91}, no. 5, 055007 (2015)
  [arXiv:1411.3310 [hep-ph]].

\bibitem{Craig:2015pha} 
  N.~Craig, A.~Katz, M.~Strassler and R.~Sundrum,
  arXiv:1501.05310 [hep-ph].


\bibitem{newexperiments} 
 P.~W.~Graham,  D.~Kaplan, J.~Mardon and S.~ Rajendran, to appear
 
 

 


\bibitem{Giudice:2008bi} 
  G.~F.~Giudice,
  In *Kane, Gordon (ed.), Pierce, Aaron (ed.): Perspectives on LHC physics* 155-178
  [arXiv:0801.2562 [hep-ph]].

\end{thebibliography}
\end{document}